\documentclass{PoS}

\title{The EFT and FCNC interpretations in the processes with top quarks at CMS}

\ShortTitle{}

\author{\speaker{K. Skovpen (On behalf of the CMS Collaboration)}\\
        Vrije Universiteit Brussel, IIHE, Pleinlaan 2, 1050 Brussels,
	Belgium\\
        E-mail: \email{kirill.skovpen@cern.ch}}

\abstract{In absence of any distinct evidence of new physics
phenomena at the LHC, an increasing number of experimental studies aim at
probing anomalous effects with an effective field theory (EFT) that represents a comprehensive
approach for interpretation of various experimental results. The
processes with the production of top quarks are sensitive to several
classes of EFT operators including the flavour-changing neutral
currents (FCNC). The summary of the latest CMS results based on the recent
studies of the standard model processes and searches for new physics effects 
involving top quarks are presented.}

\FullConference{
European Physical Society Conference on High Energy Physics - EPS-HEP2019 -\\
			10-17 July, 2019\\
			Ghent, Belgium}

\begin{document}

\section{Introduction}

The standard model (SM) is incomplete. Nevertheless, we have not found
any experimental evidence of new physics yet. It is possible that new
particles that are considered in many beyond the SM (BSM) theories
have masses that are above the energy reach at the Large Hadron Collider (LHC). The
presence of new heavy particles can potentially induce various
anomalous interactions at the electroweak and these effects can be
studied at the LHC. Possible deviations from
the SM predictions can be parametrized in a general way with an
effective field theory (EFT) approach that uses an extended SM lagrangian
built on dimension-six operators (SMEFT)~\cite{PUB_EFT1,PUB_EFT2}. 
A full classification of the EFT operators relevant to the
processes with top quarks is given in Ref.~\cite{PUB_SMEFT}.

The so-called flavour-changing neutral currents (FCNC) are forbidden
at tree level in the SM. These processes can only occur at higher
orders and are highly suppressed due to the Glashow-Iliopoupos-Maiani
(GIM) mechanism~\cite{PUB_GIM}. However, the FCNC transitions can be
significantly enhanced by several orders of magnitude in various BSM scenarios~\cite{PUB_FCNC}. 
The FCNC interactions with top quarks are particularly suppressed in the
SM with the branching fractions of the top quark FCNC decays predicted at
the level of $10^{-17}-10^{-12}$. In some of the best motivated BSM
scenarios, such as the Two Higgs Doublet Model and the Minimal
Supersymmetric Standard Model, these probabilities can be increased up to
$10^{-10}-10^{-3}$. Such high rates can be probed in the ongoing
experiments at the LHC, representing an excellent probe of BSM phenomena via
a FCNC study.

\section{Analysis of dilepton events}

The processes with the production of a top quark in association with a W boson ($\mathrm{tW}$) 
are associated with the final states containing two opposite-sign
leptons. These final states also receive a significant contribution
from the production of top quark pairs ($\mathrm{t\bar{t}}$). A simultaneous study of the $\mathrm{tW}$ and $\mathrm{t\bar{t}}$
processes in dilepton events was done at CMS~\cite{PUB_CMS} using
36~$\mathrm{fb^{-1}}$ of 13~TeV data~\cite{PUB_2L}. This is the first
study to directly constrain EFT in data using $\mathrm{tW}$ events.
Both the $\mathrm{tW}$ and $\mathrm{t\bar{t}}$ processes are sensitive to
the chromomagnetic dipole moments ($\mathcal{O}_{\mathrm{tG}}$).
The $\mathrm{tW}$ production is also particularly sensitive to
several EFT operators, such as the Wtb ($\mathcal{O}_{\mathrm{\phi q}}^{(3)}$ and
$\mathcal{O}_{\mathrm{tW}}$) and the FCNC ($\mathcal{O}_{\mathrm{uG}}$ and
$\mathcal{O}_{\mathrm{cG}}$) couplings at the production.
The triple gluon effective couplings ($\mathcal{O}_{\mathrm{G}}$)
contribute to the $\mathrm{t\bar{t}}$ production. For the
effective couplings not involving FCNC interactions, any possible deviations from the SM
predictions are dominated by the interference term between the SM and new
physics diagrams, which is linear with respect to an EFT operator. 
For small values of the non-FCNC effective couplings the predicted
distributions approach those of the SM processes, and therefore only
the production rates of the $\mathrm{tW}$ and $\mathrm{t\bar{t}}$
processes are used to derive the limits on the EFT Wilson coefficients. 
A multivariate analysis (MVA) approach is used to distinguish
between the $\mathrm{tW}$ and $\mathrm{t\bar{t}}$ production processes, as well
as to select FCNC events. One of the important systematic
uncertainties in this analysis is associated with theoretical
prediction of the studied processes. The measured EFT constraints 
on the Wilson coefficients are consistent with the SM predictions.

\begin{figure}[hbtp]
\begin{center}
\includegraphics[width=0.65\linewidth]{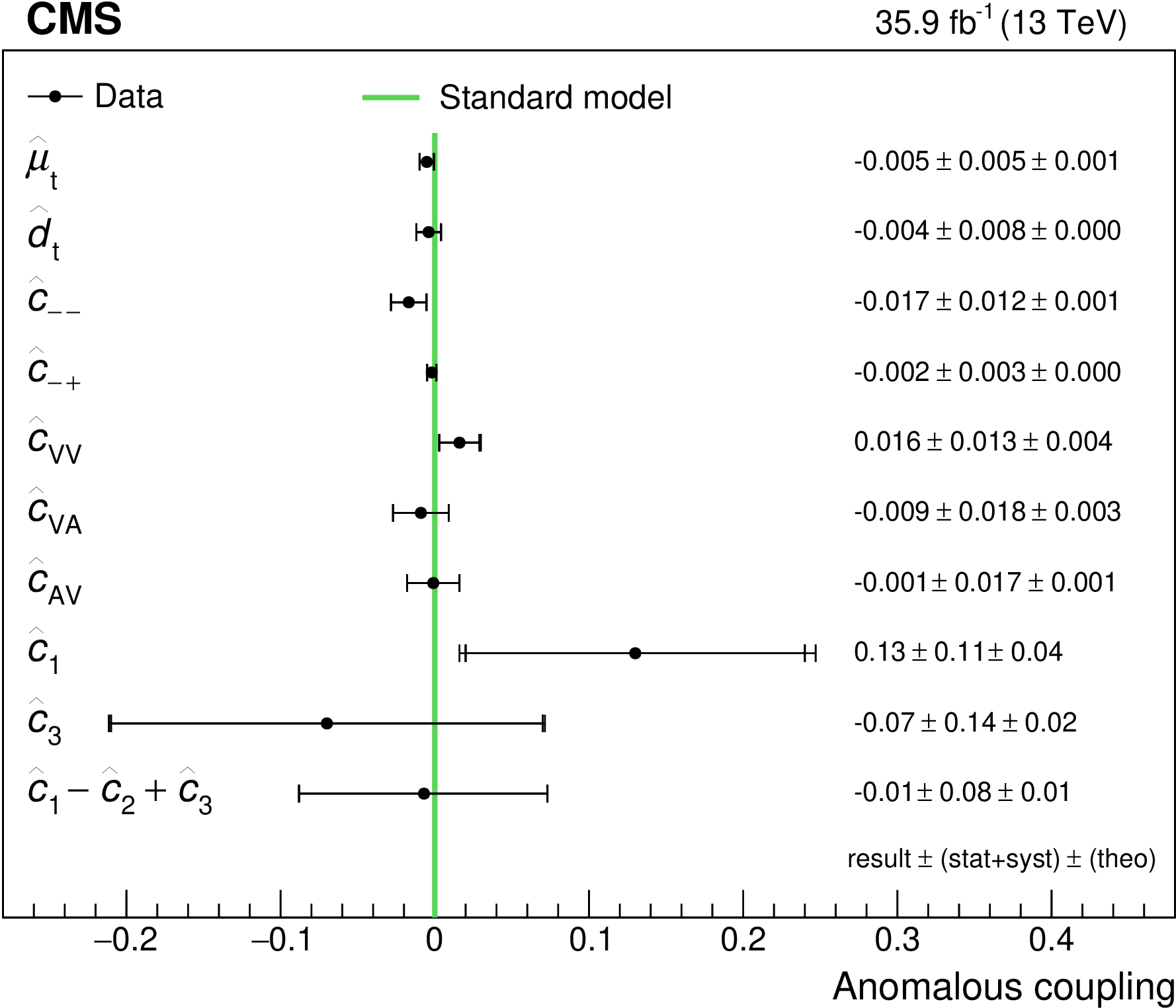}
\caption{The measured constraints on the anomalous couplings of effective
operators, including the chromomagnetic ($\hat{\mu}_t$) and
choromoelectric ($\hat{d}_t$) dipole moments~\cite{PUB_CDM}.
One non-vanishing Wilson coefficient is assumed at a time.
}
\label{fig:CDM}
\end{center}
\end{figure}

A study of spin correlations and differential cross sections of the
$\mathrm{t\bar{t}}$ production is sensitive to the top
quark chromomagnetic (CMDM, $\mathcal{O}_{\mathrm{tG}}$) and
chromoelectric dipole moments (CEDM,
$\mathcal{O}_{\mathrm{tG}}^{\mathrm{I}}$). The analysis
of 36~$\mathrm{fb^{-1}}$ of 13~TeV CMS data focuses on the study
of the properties of the spin density matrix as a function of the partonic
initial state and production kinematics~\cite{PUB_CDM}. The density matrix is
decomposed in a series of coefficients using the Pauli matrix basis to
define experimental observables sensitive to the CMDM and CEDM.
Due to the measurement of the full density matrix, the sensitivity to
the CMDM and CEDM has been significantly
enhanced with respect to the previously published direct constraints
on these interactions. The modelling uncertainties associated with the
simulation of the studied processes are among the dominant
contributions to the total systematic uncertainty in this analysis.
The resultant constraints on the studied anomalous effective
couplings, including the four-quark and two-quark-gluon(s) operators, are
presented in Fig.~\ref{fig:CDM}. The results also include
two-dimensional limits on effective couplings.

\section{Search for the four top quark production}

The process with the production of four top quarks
($\mathrm{t\bar{t}t\bar{t}}$) provides an important test
of the QCD predictions and is also sensitive to the four-fermion EFT operators
($\mathcal{O}_{\mathrm{tt}}^{\mathrm{1}}$,
$\mathcal{O}_{\mathrm{QQ}}^{\mathrm{1}}$,
$\mathcal{O}_{\mathrm{Qt}}^{\mathrm{1}}$, and
$\mathcal{O}_{\mathrm{Qt}}^{\mathrm{8}}$). The
study of this process uses 36~$\mathrm{fb^{-1}}$ of 13~TeV CMS data
and is performed in a combination of several search
channels: same-sign and opposite-sign dilepton, as well as single
lepton and trilepton final states~\cite{PUB_4T}. Potential new physics
effects are probed in a restricted BSM scenario where it is assumed
that new physics couples predominantly to the
left-handed third generation quark doublet and the right-handed top
quark singlet. The analysis uses the reconstruction of the hadronic
top quark decays and defines several event categories based on the
number of reconstructed jets. An MVA approach is used to suppress relevant background
processes. The dominant uncertainties in the analysis are
associated with the limited statistics in data, heavy-flavour jet
identification, as well as jet reconstruction.
The exclusion intervals on the EFT Wilson coefficients
are extracted from the upper limit on the $\mathrm{t\bar{t}t\bar{t}}$ production cross
section. The resultant constraints are consistent with the SM
predictions. The observed (expected) sensitivity to the
$\mathrm{t\bar{t}t\bar{t}}$
production was obtained at the level of 1.4 (1.1) standard deviations.

\section{Study of the associated production of top quarks with vector bosons}

The study of the ttbar production in association with a $\mathrm{W}$
($\mathrm{t\bar{t}W}$) or a $\mathrm{Z}$ ($\mathrm{t\bar{t}Z}$) boson is particularly sensitive 
to the electroweak couplings of the top quark. These
production processes also represent an important background in the
study of the Higgs boson ($\mathrm{H}$) production in association with top quark
pairs ($\mathrm{t\bar{t}H}$). An EFT study that is performed at CMS with 36~$\mathrm{fb^{-1}}$ of
13~TeV data considers the $\mathrm{t\bar{t}W}$ and $\mathrm{t\bar{t}Z}$ production processes
together with the contributions from the $\mathrm{t\bar{t}H}$ production~\cite{PUB_TTV}. From a large number of 
EFT operators potentially sensitive to the studied processes, only those are kept
that are exclusively associated with an enhanced sensitivity to the
$\mathrm{t\bar{t}W}$ and $\mathrm{t\bar{t}Z}$ productions.
Eight EFT operators were selected that are associated with the top quark
electroweak couplings, as well as triple gluon field and the Higgs
boson couplings. The analysis is done in the
same-sign dilepton, trilepton and four-lepton final states. Events are classified based
on the number of additional reconstructed jets. The limits on the
anomalous effective couplings are extracted from a combined likelihood fit over
the defined event categories. The resultant constraints are in agreement with the SM
predictions. The main systematic uncertainties are associated with the 
integrated luminosity, lepton identification, trigger selection
efficiencies, and non-prompt lepton backgrounds.

\begin{figure}[hbtp]
\begin{center}
\includegraphics[width=0.65\linewidth]{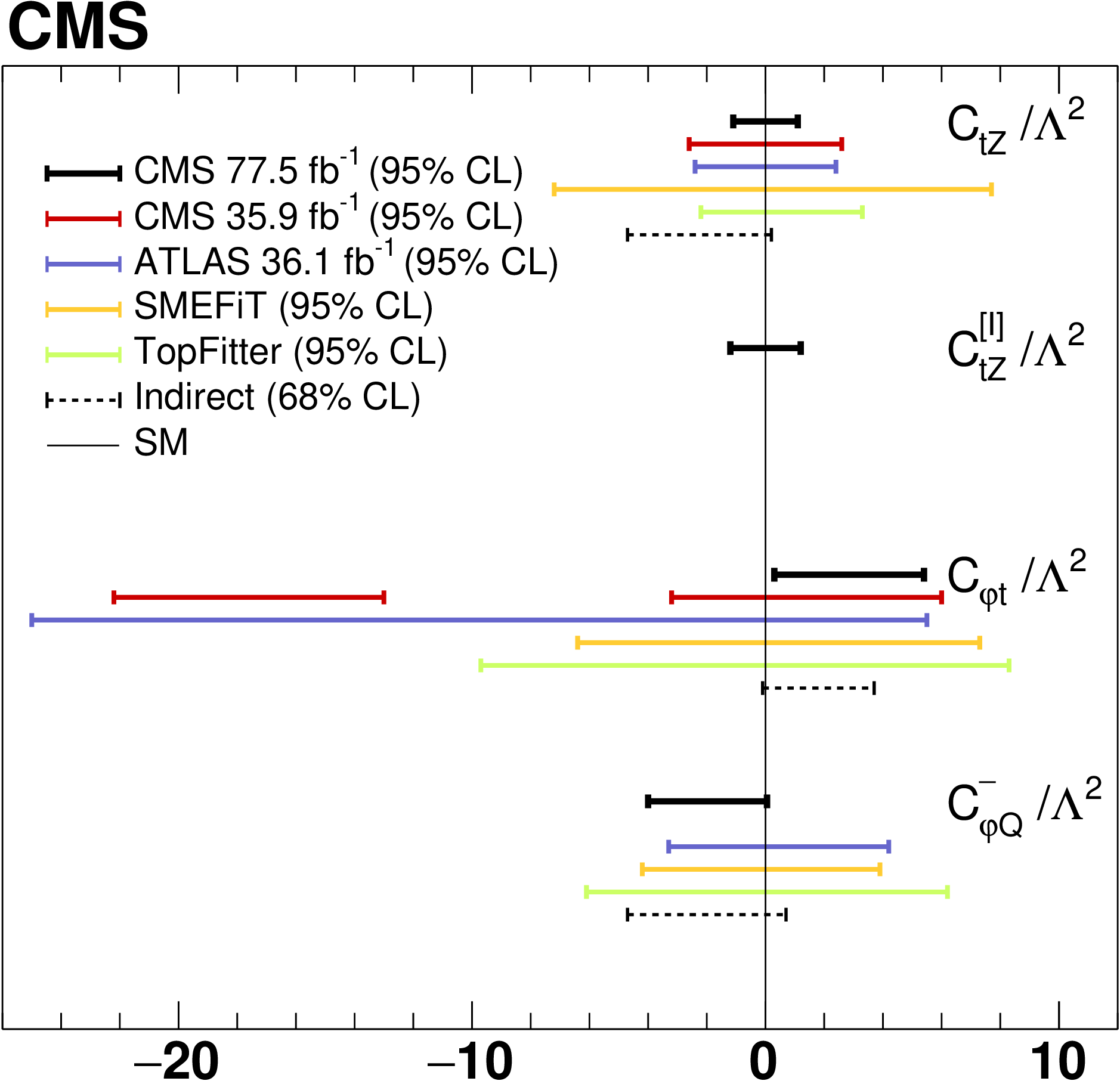}
\caption{The observed constraints on the Wilson coefficients extracted
from the differential cross section measurement of the
$\mathrm{t\bar{t}Z}$ process~\cite{PUB_TTZ}. The results are compared
to various previous experimental constraints and results of combined fits relevant to these operators.}
\label{fig:TTZ}
\end{center}
\end{figure}

The measurement of the differential $\mathrm{t\bar{t}Z}$ production
cross section is done at CMS using 78~$\mathrm{fb^{-1}}$ of 13 TeV data~\cite{PUB_TTZ}. 
The analysis considers final states with three
and four leptons. The EFT operators relevant to this study are the electroweak
dipole moments ($\mathcal{O}_{\mathrm{tZ}}$, $\mathcal{O}_{\mathrm{tZ}}^{\mathrm{[I]}}$) and anomalous
neutral-current interactions ($\mathcal{O}_{\mathrm{\phi t}}$,
$\mathcal{O}_{\mathrm{\phi Q}}^{-}$).
The analysis strategy and the dominant sources of systematic uncertainties are
similar to Ref.~\cite{PUB_TTV}. The study uses a reweighting approach
that is applied to the generator-level information to predict BSM effects at
next-to-leading order with a proper treatment of systematic correlations.
The statistical uncertainty is
comparable to the total systematic uncertainty in this measurement.
In addition to the EFT study, this analysis includes constraints on
the vector and axial-vector current couplings, as well as the electroweak
dipole moments. The final results are presented as one- and
two-dimensional constraints on the effective couplings. A comparison
of the resultant one-dimensional constraints on the Wilson coefficients 
to various direct and indirect studies is shown in Fig.~\ref{fig:TTZ}.

\section{Search for FCNC with top quarks}

\begin{figure}[hbtp]
\begin{center}
\includegraphics[width=0.75\linewidth]{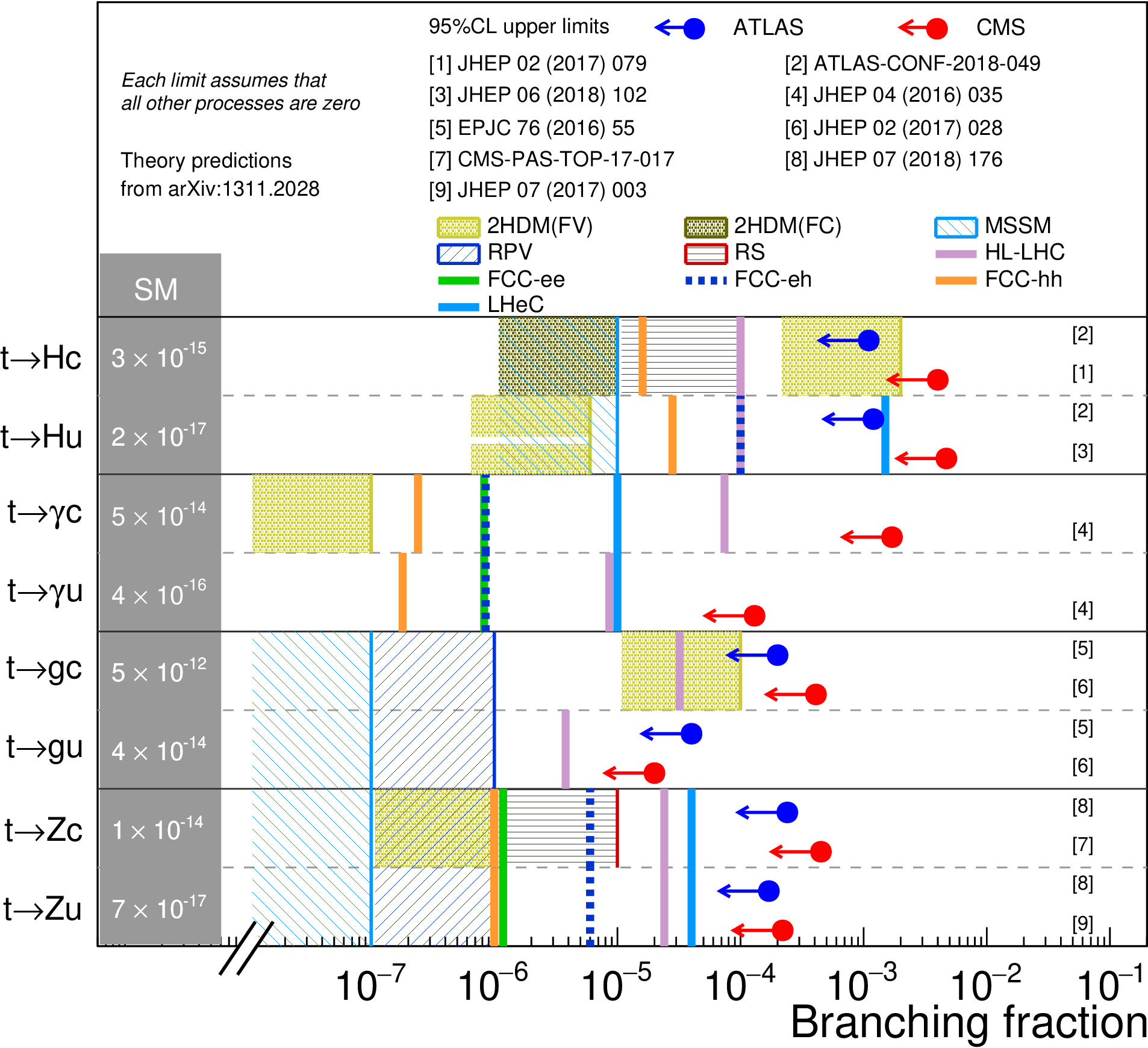}
\caption{The present results and future projections for the limits
on the top quark FCNC decay branching fractions~\cite{PUB_FCNC2}.
The experimental constraints are compared to various BSM predictions~\cite{PUB_FCNC}.}
\label{fig:FCNC}
\end{center}
\end{figure}

The FCNC processes with top quarks are studied in various search channels
at CMS. The summary of the latest results is presented in
Fig.~\ref{fig:FCNC}.

A search for the top quark FCNC interactions with the Higgs
boson is done for the case of the $\mathrm{H \to b\bar{b}}$
decays using 36~$\mathrm{fb^{-1}}$ of 13 TeV data~\cite{PUB_FCNC3}.
This study considers top FCNC interactions in events with the production of top quark pairs, 
as well as in the associated production of a top quark and a Higgs boson. The latter is
only sensitive to the top FCNC interaction with an up quark that
is enhanced due to the proton parton distribution function. The
analysis introduces several event categories defined by the number of
reconstructed jets originating from an hadronization process of b quarks. The main
systematic uncertainty in this analysis is associated with the
identification of heavy flavour jets. The extraction of
final constraints on the top FCNC couplings is performed with a
likelihood fit over the defined event categories. The observed
(expected) contraints on the $\mathrm{t \to Hu}$ and $\mathrm{t \to
Hc}$ branching fractions are 0.47\% (0.34\%) and 0.47\% (0.44\%),
respectively.

A search for the top FCNC interactions with a Z boson is done in the
final states with three leptons and is based on the analysis of
36~$\mathrm{fb^{-1}}$ of data collected at 13 TeV~\cite{PUB_FCNC4}. The FCNC
effects are studied in the top quark decays and in the process of a top quark
associated production with a Z boson. In the study of each of the production
channels a signal kinematic region is defined using an MVA approach.
The main systematic
uncertainties are associated with non-prompt leptons and
normalization of the background processes with the production of
multiple W and Z bosons, as well as the $\mathrm{t\bar{t}Z}$
production process. The FCNC constraints are extracted from a
combined likelihood fit based on the distributions of the MVA
discriminants obtained in the signal regions. The
observed (expected) limits on the $\mathrm{t \to Zu}$ and $\mathrm{t
\to Zc}$ branching fractions are 0.024\% (0.015\%) and 0.045\% (0.037\%),
respectively.

\section{Future prospects and summary}

\begin{figure}[hbtp]
\begin{center}
\includegraphics[width=0.45\linewidth]{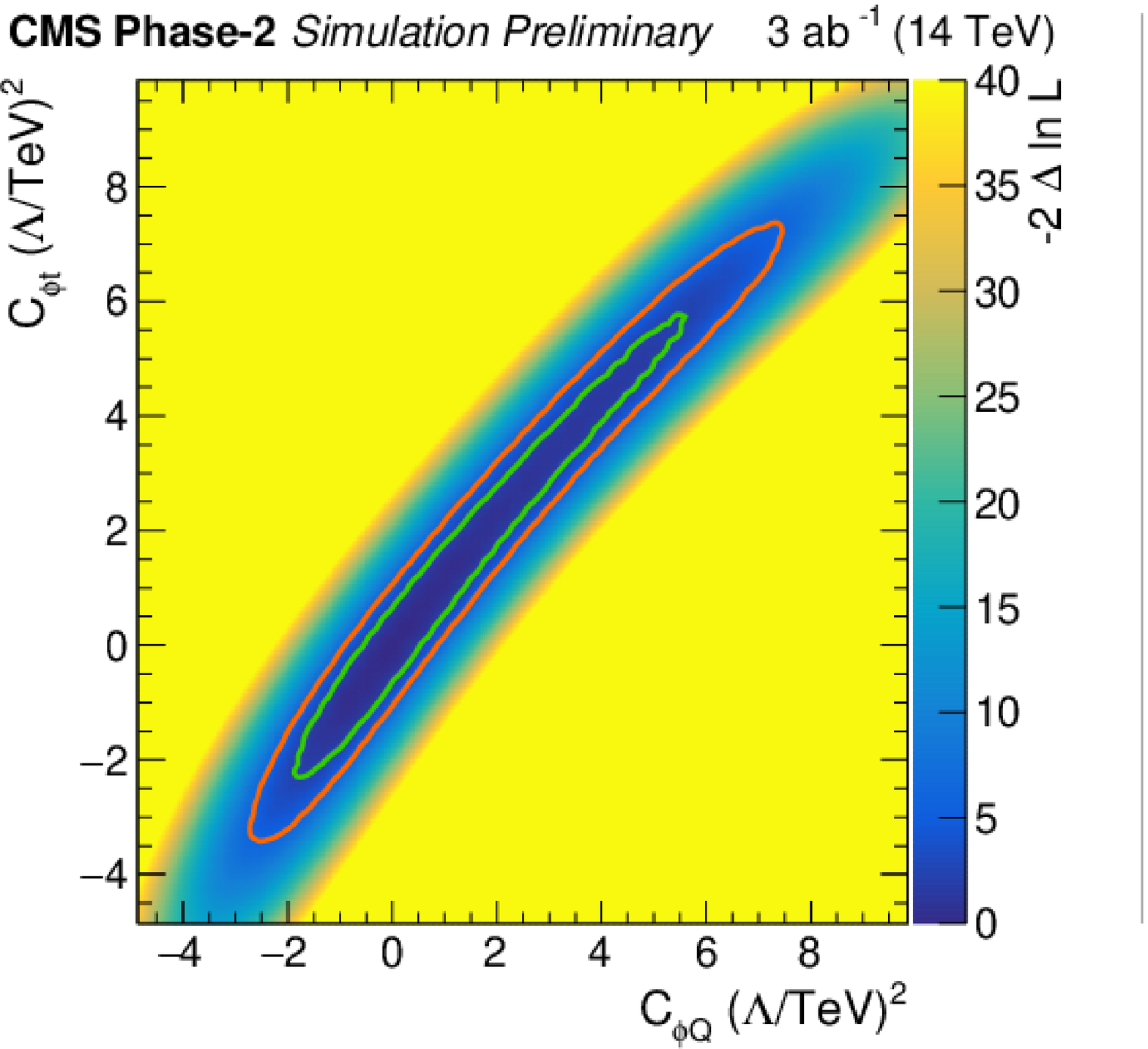}
\includegraphics[width=0.45\linewidth]{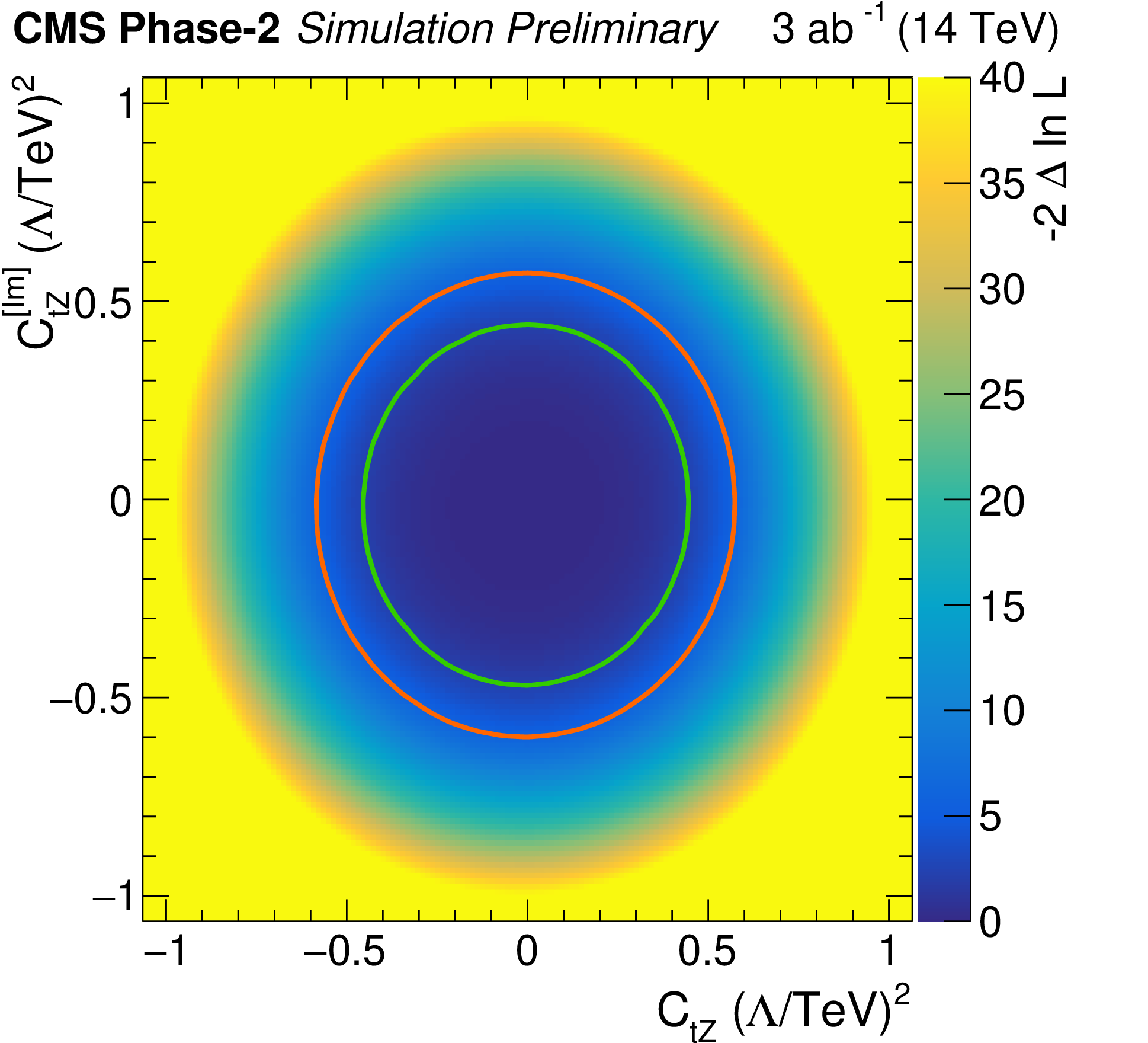}
\caption{A negative likelihood as a function of the Wilson coefficients: $\mathcal{O}_{\mathrm{\phi
Q}}^{-}$/$\mathcal{O}_{\mathrm{\phi t}}$ (left) and
$\mathcal{O}_{\mathrm{tZ}}$/$\mathcal{O}_{\mathrm{tZ}^{\mathrm{[I]}}}$
(right)~\cite{PUB_FUTSM2}. The 68\% and 95\% confidence
level contours are shown in green and red colours, respectively.}
\label{fig:PROJ}
\end{center}
\end{figure}

Future studies of the EFT anomalous couplings and FCNC
interactions in the top quark sector at the High Luminosity LHC (HL-LHC)
are associated with an important gain in the sensitivity with respect to
the present limits obtained on these interactions. A preliminary study of the $\mathrm{t\bar{t}Z}$ process at
the HL-LHC resulted in the EFT constraints that are better by a factor
of two than the current best limits obtained at the
LHC~\cite{PUB_FUTSM,PUB_FUTSM2}. The projected two-dimensional limits
on the EFT Wilson coefficients are shown in Fig.~\ref{fig:PROJ}.
A similar level of improvement in constraining EFT operators is also
anticipated in the analysis of the $\mathrm{t\bar{t}t\bar{t}}$ production~\cite{PUB_FUTSM3}. The HL-LHC
data set will allow to further constrain the anomalous top FCNC decays. 
The projected sensitivities are expected to surpass by almost one order of magnitude
the current best limits~\cite{PUB_FCNC2}. The future
projections for the EFT studies of the FCNC processes with top quarks show
an improvement of a factor of two in the final constraints on the
two-fermion operators~\cite{PUB_FUTFCNC}

Over the past few years there has been an increasing number of experimental
analyses in the top quark sector using an EFT approach to probe new physics effects. These
studies provide an important set of constraints that can be easily
reinterpreted for a given BSM model. A lot of progress was
done with the full classification of the relevant EFT operators for
the processes involving top quarks, providing an important guidance for
various ongoing experimental searches at the LHC. The
future studies at the HL-LHC are expected to significantly improve the current best
limits on the EFT and FCNC couplings.

\end{document}